\documentclass[journal]{IEEEtran}

\usepackage[utf8]{inputenc}                             
\DeclareUnicodeCharacter{FB01}{fi}

\usepackage[colorlinks,linkcolor=black,citecolor=black,urlcolor=black]{hyperref}
\usepackage{cite}                                       
\usepackage[pdftex]{graphicx}                           
\usepackage[caption=false,font=footnotesize]{subfig}    
\usepackage{lipsum}
\usepackage[usenames,dvipsnames,svgnames,table]{xcolor}
\usepackage{amsmath}
\usepackage{url}
\usepackage{listings}
\usepackage{etoolbox}
\usepackage{float}
\usepackage{booktabs}
\usepackage{microtype}
\usepackage{cleveref}
\usepackage{easyReview}
\usepackage{textcomp}
\usepackage{caption}
\usepackage{graphicx}
\usepackage[inline]{enumitem}
\usepackage{comment}

\crefname{figure}{fig.}{Fig.}           

\usepackage[belowskip=-15pt,aboveskip=0pt]{caption}
 
\begin{document}


\title{End-to-end Quantum Secured Inter-Domain 5G Service Orchestration Over Dynamically Switched Flex-Grid Optical Networks Enabled by a q-ROADM}

\author{R.~Wang,
        R.~S.~Tessinari,
        E.~Hugues-Salas,
        A.~Bravalheri,
        N.~Uniyal,
        A.~S.~Muqaddas,
        R.~S.~Guimaraes,
        T.~Diallo,
        S.~Moazzeni,
        Q.~Wang,
        G.~T.~Kanellos,
        R.~Nejabati, and
        D.~Simeonidou
        \vspace{-8mm}

\thanks{R.~Wang, R.~S.~Tessinari, E.~Hugues-Salas, A.~Bravalheri, N.~Uniyal, A.~S.~Muqaddas, R.~S.~Guimaraes, T.~Diallo, S.~Moazzeni, Q.~Wang, G.~T.~Kanellos, R.~Nejabati, and D.~Simeonidou are with High Performance Network Group, Department of Electrical and Electronic Engineering, University of Bristol, UK (e-mail: rui.wang@bristol.ac.uk).}


}

\markboth{Journal of Lightwave Technology, VOL. XX, No. X, 2019.}{}

\maketitle

\begin{abstract}
Dynamic and flexible optical networking combined with virtualization and softwarisation enabled by Network Function Virtualization (NFV) and Software Defined Networking (SDN) are the key technology enablers for supporting the dynamicity, bandwidth and latency requirements of emerging 5G network services. To achieve the end-to-end connectivity objective of 5G, Network Services (NSes) must be often deployed transparently over multiple administrative and technological domains.  Such scenario often presents security risks since a typical NS\footnote{Network Service is a combination of multiple virtual and physical network functions created to realise a desired network functionality} may comprise a chain of network functions, each executed in different remote locations, and tampering within the network infrastructure may compromise their communication. To avoid such threats, Quantum Key Distribution (QKD) has been identified and proposed as a future-proof method immune to any algorithmic cryptanalysis based on fundamental quantum-physics mechanisms to distribute symmetric keys. The maturity of QKD has enabled the research and development of quantum networks with gradual coexistence with classical optical networks using carrier-grade telecom equipment. This makes the QKD technology a suitable candidate for security of distributed and virtualised  network services.

In this paper, for the first time, we propose a dynamic quantum-secured optical network for supporting network services that are dynamically created by chaining Virtual Network Functions (VNFs\footnote{Hardware network functions when implemented in software and deployed as VMs or containers are called VNFs}) over multiple network domains.  This work includes a new flex-grid quantum-switched Reconfigurable Optical Add Drop Multiplexer (q-ROADM), extensions to SDN-enabled optical control plane, and extensions to NFV orchestration to achieve quantum-aware, on-demand chaining of VNFs. The experimental results verify the capability of routing quantum and classical data channels both individually and dynamically over shared fibre links. Moreover, quantum secured chaining of VNFs in 5G networks is experimentally demonstrated via interconnecting four autonomous 5G islands simultaneously through the q-ROADM with eight optical channels using the 5GUK Exchange orchestration platform. The experimental scenarios and results confirm the benefit of the proposed data plane architecture and control/management plane framework.

\end{abstract}


\IEEEpeerreviewmaketitle

\section{Introduction}
With the mature techniques for radio access, fronthaul, and backhaul networks available, the evolving 5G architecture needs to focus on developing solutions for managing disaggregation and cloudification of network functions and guarantee the security of end-to-end 5G network services \cite{akyildiz2015softair, wubben2014benefits}. Managing disaggregation and cloudification at large scale is fundamental to future 5G service-offering, as it enables new types of 5G services including cross-domain network slicing and advanced neutral hosting scenarios, therefore delivering the real value promised by 5G \cite{li2017network} \cite{rost2014cloud}. In this paper, we orchestrate services over multiple network domains with heterogeneous network devices and cloud infrastructure to deploy end-to-end services \cite{5gpppArch}.

5G networks are expected to be the driving force behind verticals such as smart cities, digital health and manufacturing enabled by meeting the stringent KPI requirements including latency, bandwidth and massive connectivity \cite{5gppp}\cite{ngmn}. Dynamic and rapid deployment of network services using techniques like SDN and Network Function Virtualization (NFV) have allowed telecom operators and service providers to meet such strict KPIs. Organisations like European Telecommunications Standards Institute (ETSI), Metro Ethernet Forum, and Open Networking Foundation have driven the development of systems such as NFV Orchestrators (NFVO), allowing service providers to create and deploy the network services on the commodity hardware as VNFs, reducing the CAPEX and OPEX as well as making the system more fault tolerant and dynamic.

However, one of the major objectives of 5G is to achieve the end-to-end connectivity over heterogeneous network segments \cite{ngmn}\cite{5gppp}, which most of the current NFVO does not support. With edge and cloud systems working coordinately, the NFVO systems have been designed towards creating a seamless interconnection in a single domain\cite{osmWIM}. As discussed in \cite{OrchSurvey}, there has been recent work done in the field of multi-domain orchestration allowing service providers to create end-to-end network services irrespective of the underlying network administrative domains. Such systems abstract the distribution and disaggregation of resources across multiple network domains for the end-users. In addition, optical networks play an essential role in connecting these services while achieving the desired KPIs \cite{osmWIM}. Despite these developments, end-to-end security remains a concern and could prevent its full operation. For instance, authors in~\cite{yang2017survey} describe different security threats over MECs, considering Internet-of-Things (IoT) devices. Security and privacy attacks are also illustrated in \cite{ni2017securing} over fog computing architectures in which off-the-shelf solutions are proposed to eliminate these threats.

Quantum key distribution is a technology to utilise quantum mechanisms to encrypt and decrypt key information. Therefore, quantum cryptography is being considered as the technology for network security recently since the quantum-grade encryption mechanism proves to be information theoretical secure. Thus it can be regarded as the ultimate technology to strengthen the physical layer communication security~\cite{pirandola2019advances}. Worldwide demonstrations confirm the practicality of quantum encryption including different testbeds in Vienna, Tokyo, Battelle, Cambridge, Florence and China (Beijing-Shanghai) \cite{peev2009secoqc,sasaki2011field,walenta2015towards,bacco2019field,wonfor2017high,mao2018integrating}. Moreover, the coexistence of the quantum channel and classical data channels prove the feasibility of quantum encryption together with standard telecommunication channels without the need of surplus optical fibre for point-to-point transmission \cite{eriksson2019wavelength, patel2014quantum, lin2018telecom}.

From the networking point of view, approaches integrating QKD and Software Defined Networking (SDN) have been reported in \cite{aguado2016quantum, aguado2017secure,zhao2018resource,hugues2019monitoring}. From these research works, it has been proved that SDN is beneficial for QKD enabled optical networks to optimise the performance of the QKD links after applying a customised network configuration. For instance, it is demonstrated~\cite{hugues2019monitoring} that real-time monitoring of quantum parameters provides information to the SDN controller to secure lightpaths in the optical network with flexible optical path configurations to ensure the uninterrupted distribution of quantum keys in case of physical layer attacks. Furthermore, the use of QKD for NFV has also been studied and several demonstrations verify the application of AES encryption using quantum keys to secure virtual network functions (VNFs) \cite{aguado2017secure, aguado2018virtual}. Based on these advances on QKD technologies and field trials, and their combination with SDN and NFV, quantum key distribution can also be applied to secure interconnections of distributed VNFs within 5G framework.

In this paper, we address the problem by experimentally demonstrating a quantum-secured multi-domain 5G network using an orchestrator, the 5GUK Exchange, for end-to-end service composition with the following novel features: \begin{enumerate*}[label=\arabic*), ref=\arabic*]
\item On-demand composition of Network Services (NSes) by chaining distributed VNFs hosted in data centers belonging to different 5G network domains;
\item Extending SDN control plane and OSM orchestrator to support quantum secured NFV chaining;
\item Utilising QKD for quantum-securing the NSes via a quantum meshed network enabled by a novel 4 degree quantum switched flex-grid q-ROADM; and
\item On-demand optimisation of the service quality of NSes (bandwidth and connectivity) by dynamically adjusting the optical parameters such as modulation format and power
\end{enumerate*}. We extend our previous work in \cite{nejabati2019first} by providing detailed description, implementation and more results of the proposed network architecture, {in data plane and control/management plane sides}. The experimental results validate the value of the proposed scheme. The rest of the paper is organised as follows: \Cref{sec:scenarios} describes the motivation, the concept and the scenarios of the proposed multi-domain secured network service orchestration. \Cref{sec:q-roadm} describes the architecture of the proposed q-ROADM to allow quantum channel and data channel switching. \Cref{sec:Island_connectivity} explains the experimental testbed physical connectivity and setup. \Cref{Sec:5GExArch} covers the description of control plane and management plane of the physical experimental testbed. \Cref{sec:results} reveals the data plane results and control/management results from the experiment and \Cref{sec:conclusion} concludes the paper.

\section{Concept for Multi-Domain Secure Network Services Composition} 
\label{sec:scenarios}


\begin{figure}
    \centering
    \includegraphics[width=\linewidth,trim={0 0 220 0},clip]{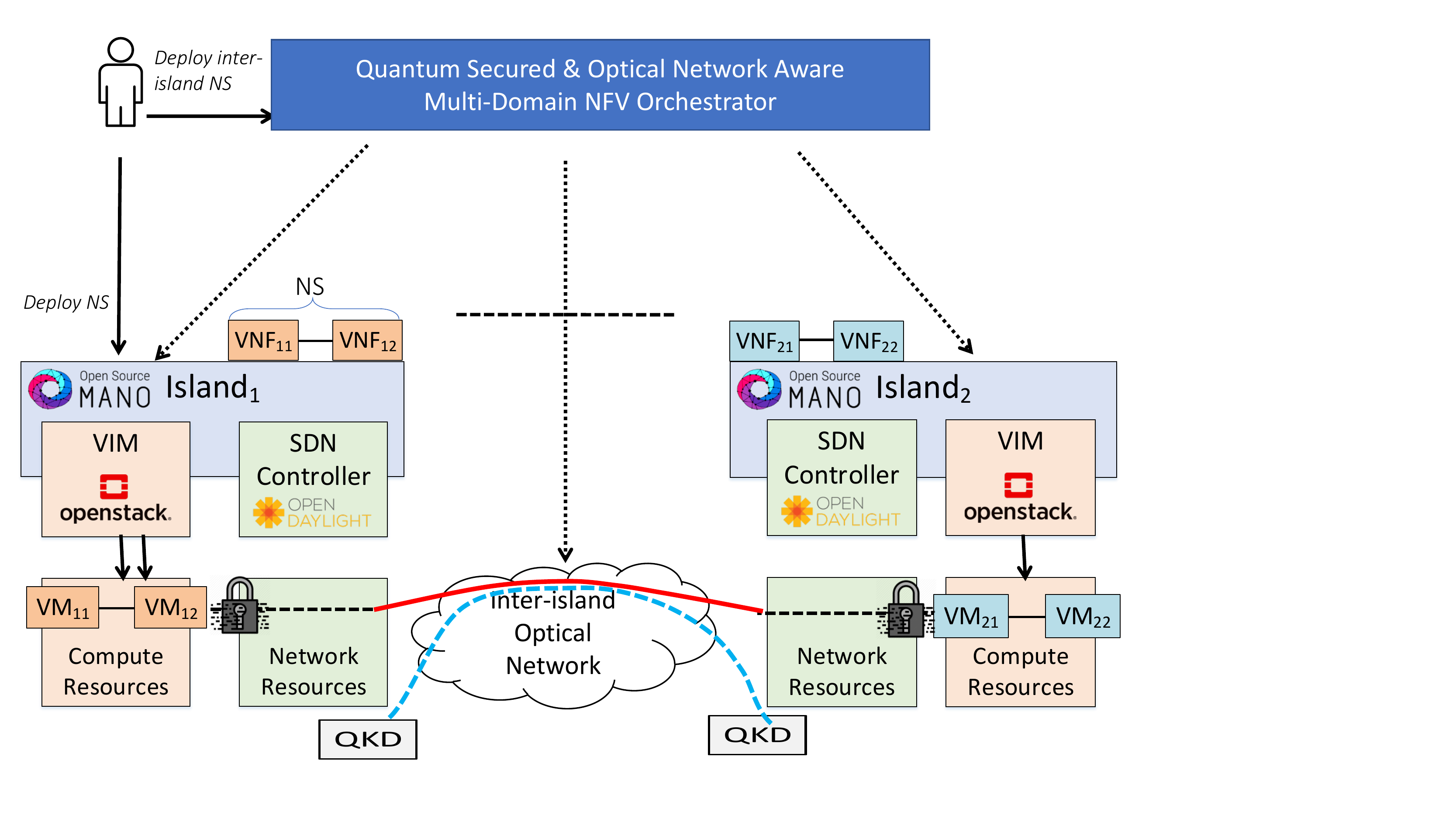}
    \caption{Overall concept of QKD secured 5GUK Exchange scenario}
    \label{fig:Overall_Description}
\end{figure}

Fig. \ref{fig:Overall_Description} depicts two individual autonomous network domains (islands) where a local NFVO provisions and controls the computing and network resources through the the Virtual Infrastructure Manager (VIM) and SDN controller. As described earlier, each NFVO could deploy and manage the NSes on a single network domain. To achieve the goal of end-to-end orchestration, we have used and extended the 5GUK Exchange \cite{5GUKEx} to create inter-island NSes. Additionally, the inter-island NSes should exhibit enhanced dynamic characteristics, provide the bandwidth and latency requirements between VNFs and effectively address security concerns raised by the distributed nature of the inter-island NSes.

To realise this, in our proposed implementation 
\begin{enumerate}
    \item We used the 5GUK Exchange \cite{5GUKEx} to support the inter-island NS management and orchestration in order to allow for mixing and matching VNFs from multiple islands
    \item We developed a new quantum switch enabled q-ROADM to provide dynamic inter-island optical network connectivity for high bandwidth and low latency VNF chaining across the multiple domains and to provide dynamical quantum key switching.
    \item We extended the use of the dynamic optical network to accommodate for quantum channels in a coexisting form, in order to create secured inter-island NS using QKD over the same fiber as classical traffic.
    \item We extended the 5GUK Exchange capabilities to provision and control the inter-island optical network resources and the QKD resources across the islands (to be discussed in \cref{Sec:5GExArch})
\end{enumerate}

To verify the complexity and dynamicity of our concept, we executed three different scenarios (a to c) in increasing order of complexity, which is shown below in \cref{sec:res:Control_data}.
{ For simplicity, we consider each island exposing multiple NSes, each consisting of a single VNF.}
To begin, we consider a scenario where an end-user wants to deploy multiple inter-island NSes between the 4 autonomous islands, known as Island 1, Island 2, Island 3, and Island 4. {Each island represents a metro network connecting 5G access network and feeding corresponding 5G traffic into core optical network through a layer 2 switch.} In scenario 1, only one communication channel between the two islands (Island 2 and Island 4) needs to be secured while other channels remain unsecured. Moving on to the next case all islands communicate with the NS between Island 1 and Island 3 also being secured along with the existing Island 2 and Island 4. Finally, to verify the dynamic nature of the VNF deployment and NS creation, we switch the quantum channels in scenario 3 (Island 1 - Island 4 and Island 2 - Island 3) with respect to scenario 2 (Island 1 - Island 3 and Island 2 - Island 4 secured). In this scenario, the 5GUK Exchange reconfigures the L0 network to switch the quantum channel while keeping the deployed NS intact, creating the minimum service disruption. During the experiment, while creating the secure channel, the respective NS wavelength is combined with an additional QKD channel and are transmitted in parallel across the optical network. For each optical and quantum channel, the wavelengths ($\lambda_1$) are communicated to the individual islands by the 5GUK Exchange which creates the L0 and L2 network between the selected islands.



\section{Q-ROADM Description} \label{sec:q-roadm}

To enable classical data channels and QKD signal routing and switching functionality, the q-ROADM needs to provide low loss switching capability for the QKD channel as QKD transmitter Alice emit single photon level light thus it is critically sensitive to power loss \cite{diamanti2016practical}. The conventional ROADMs designed for switching classical data channels, usually associated with 12 - 20 dB insertion loss \cite{tibuleac2010transmission}, is not feasible for QKD signal routing and switching. Apart from high insertion loss, the ASE noise generated from pre-amplifiers and post-amplifiers of the ROADM would fall into the quantum channel, posing a significant challenge to the QKD receiver Bob \cite{chapuran2009optical}. 

\begin{figure}
    \centering
    \includegraphics[width=\linewidth,trim={0 0 425 0},clip]{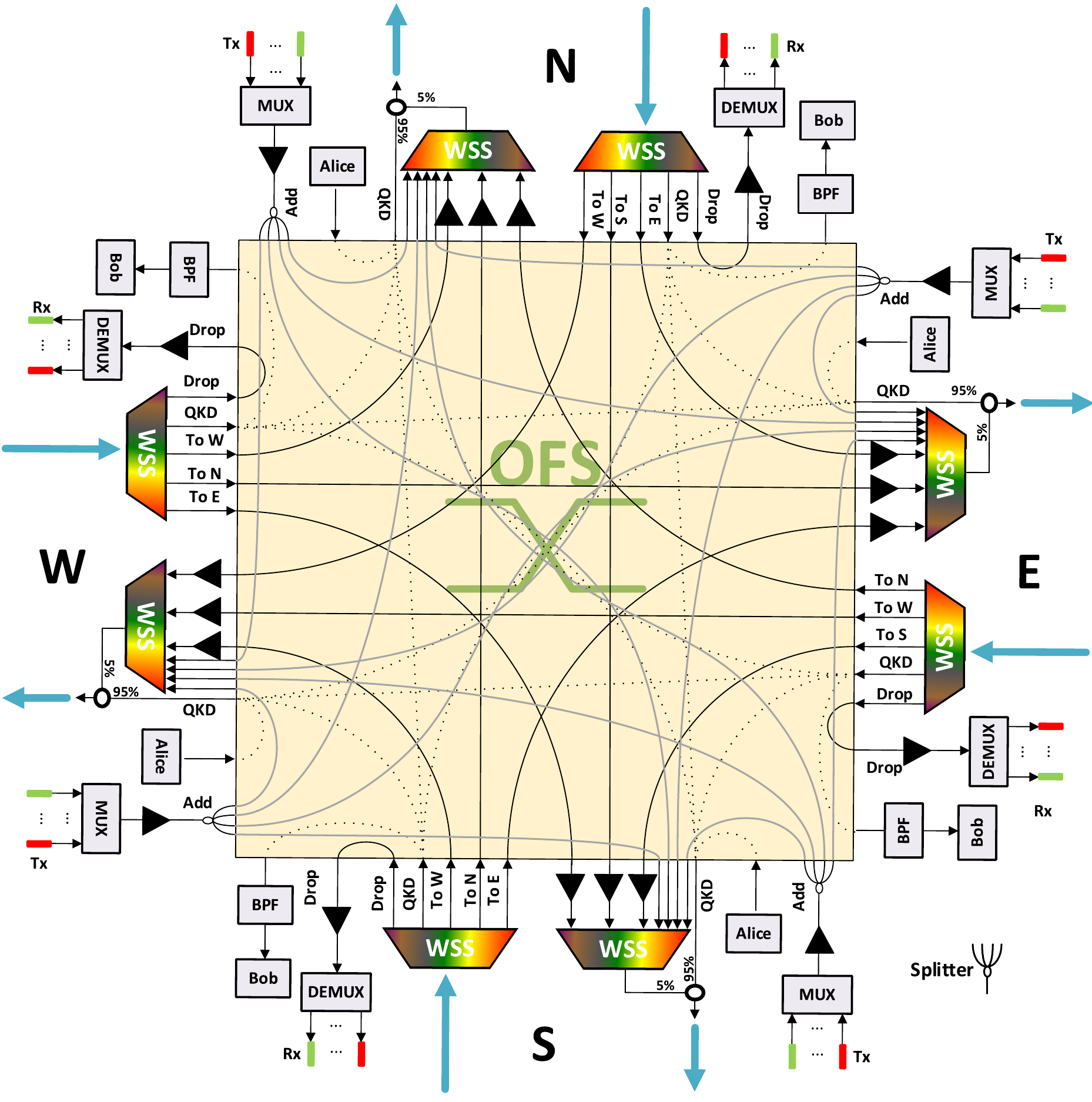}
    \caption{4 degree {CD} q-ROADM architecture design.}
    \label{fig:roadm_Arc}
\end{figure}

To overcome the drawbacks of the conventional ROADM, we propose a {colourless and directionless (CD)} flex-grid QKD enabled reconfigurable optical add-drop multiplexer architecture design based on the concept of Architecture on Demand (AoD) \cite{amaya2011architecture}, known as q-ROADM.  In \Cref{fig:roadm_Arc}, the design of a four-degree bidirectional {CD} q-ROADM is illustrated. All the optical devices are connected to the optical fibre switch (OFS) backplane, where new devices can be added to support new functions required by new network service. The AoD based q-ROADM can also be quickly programmed to increase or decrease the number of degrees as well as transform to a hybrid q-ROADM where part of degrees support coexistent quantum and data channel switching while the rest only support data channel switching.

\begin{figure*}[!t]
\centering
\includegraphics[width=\linewidth,trim={0 0 190 0},clip]{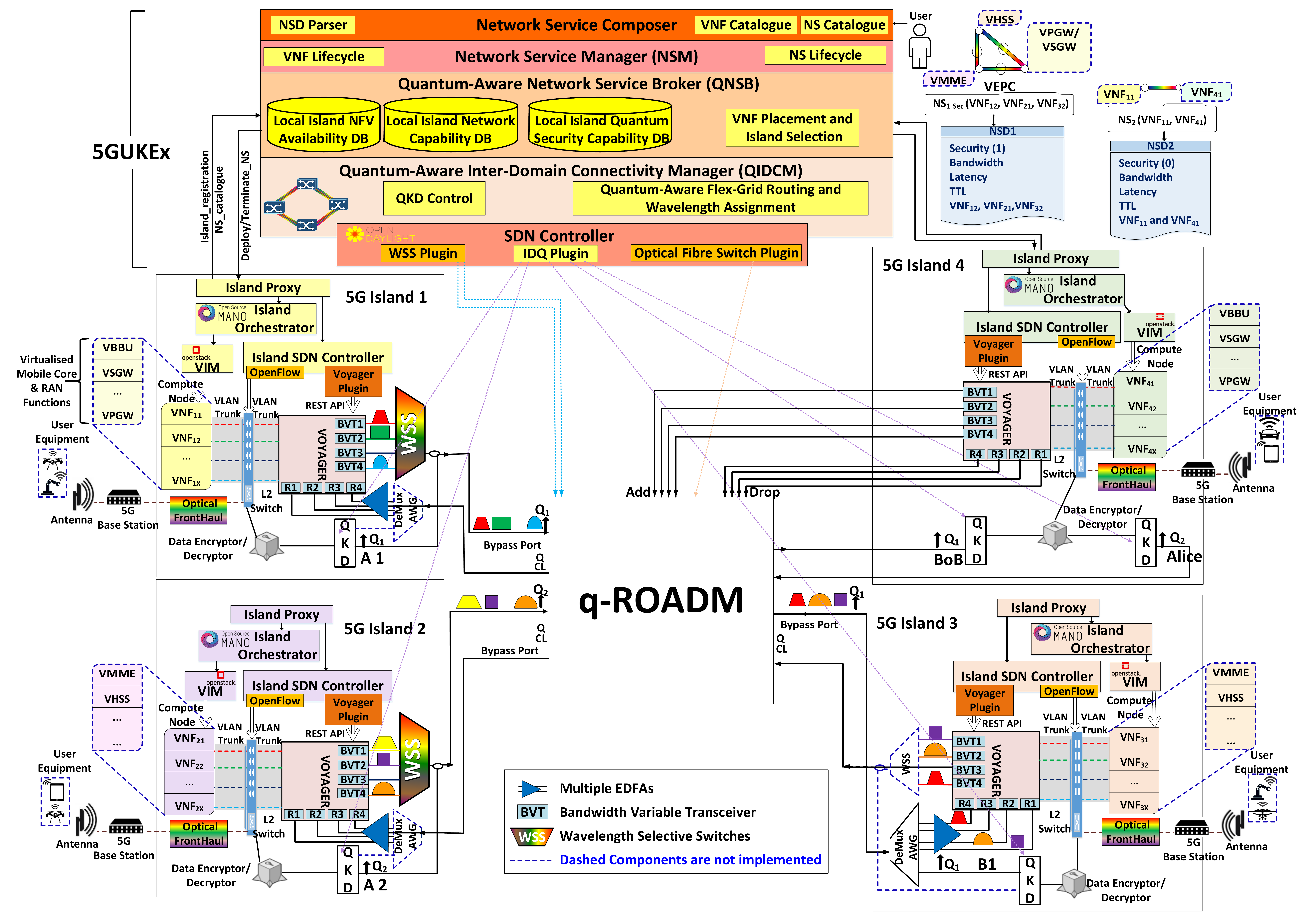}
\caption{The architecture of QKD secured 5G exchange experimental testbed setup.}
\label{fig:arch}
\end{figure*}

For each degree, the MUX/DEMUX is designed to add/drop local classic data channels, whereas the Alice or Bob is also used for adding or dropping the QKD channel locally to secure the data channel originating or terminating at the local node. The band-pass filter (BPF) placed before the quantum receiver Bob can filter out-band crosstalk from the WSS. The proposed q-ROADM architecture supports QKD and classic data channels switching functionality between any pair of ports. The QKD signal dropping and routing functionalities are realised by controlling the optical fibre switch behaviour, which is shown as the dashed line in \Cref{fig:roadm_Arc}. The only limitation is that two QKD channels cannot be routed to the same output port, due to OFS constraint as well as the colour constraint. The pre-amplifier is completely removed to avoid the ASE noise falling into the QKD channel. The output signal of post-amplifiers is combined and filtered by the WSS to avoid the ASE noise falling into the QKD channel. Moreover, the WSS at the output of each port equalises the power of different data channels and controls the total data channel power coexisting with the QKD channel of the following fibre link. The 95:5 fibre splitter is used as a low loss combiner for QKD signal to aggregate the QKD channel and the data channels. The proposed q-ROADM architecture gives around 5.3 dB loss for the QKD routing and switching, approximate 5.9 dB, and 1.2 dB power loss for dropping and adding QKD channel respectively. 

In contrast to the low loss characteristics of a quantum channel, the proposed q-ROADM architecture would introduce higher penalty to the data channels compared to the conventional ROADM as it poses higher power loss to the classical data channels while only permits post-amplifier. The power loss is around 23 dB for bypassing data channels, 21.5 dB for adding data channels, and 8.5 dB for dropping data channels. Its nature is similar to the coexistence of classical data channels and QKD channel in the fibre link only allowing extremely low power of the data channels, where the low power data channels perform worse than usual power data channels. The design of q-ROADM sacrifices the performance of the data channels, but can provide the low loss capability of adding, dropping, and switching the QKD channel. The impact of high signal quality degradation from q-ROADM to the classical data channels can be balanced by deploying robust digital signal processing techniques, such as modulation format adaption, higher forward error correction (FEC) code rate and probabilistic shaping. 


\section{Island Physical Network Description} \label{sec:Island_connectivity}

The paper implements, for the first time, a quantum-switched optical network that allows: 
\begin{enumerate*}[label=\arabic*), ref=\arabic*]
\item to dynamically control and assign wavelength, FEC coding rate, modulation format and power of each optical channel interconnecting the VNFs;
\item a coexistent scheme of quantum and classical channels where any combination of mixed or independent quantum and classical signals arriving from one island can be forwarded to any other island providing a full 4-degree q-ROADM functionality that seamlessly interconnects the four islands, {as shown in \cref{fig:arch}}; 
\item dynamic configuration of different pairs of discrete-variable (DV) QKD transmitters (Alices) and receivers (Bobs) enabling provisioning/creation/use of new quantum paths. 
\end{enumerate*}

\begin{figure*}[!t]
\centering
\includegraphics[width=\linewidth]{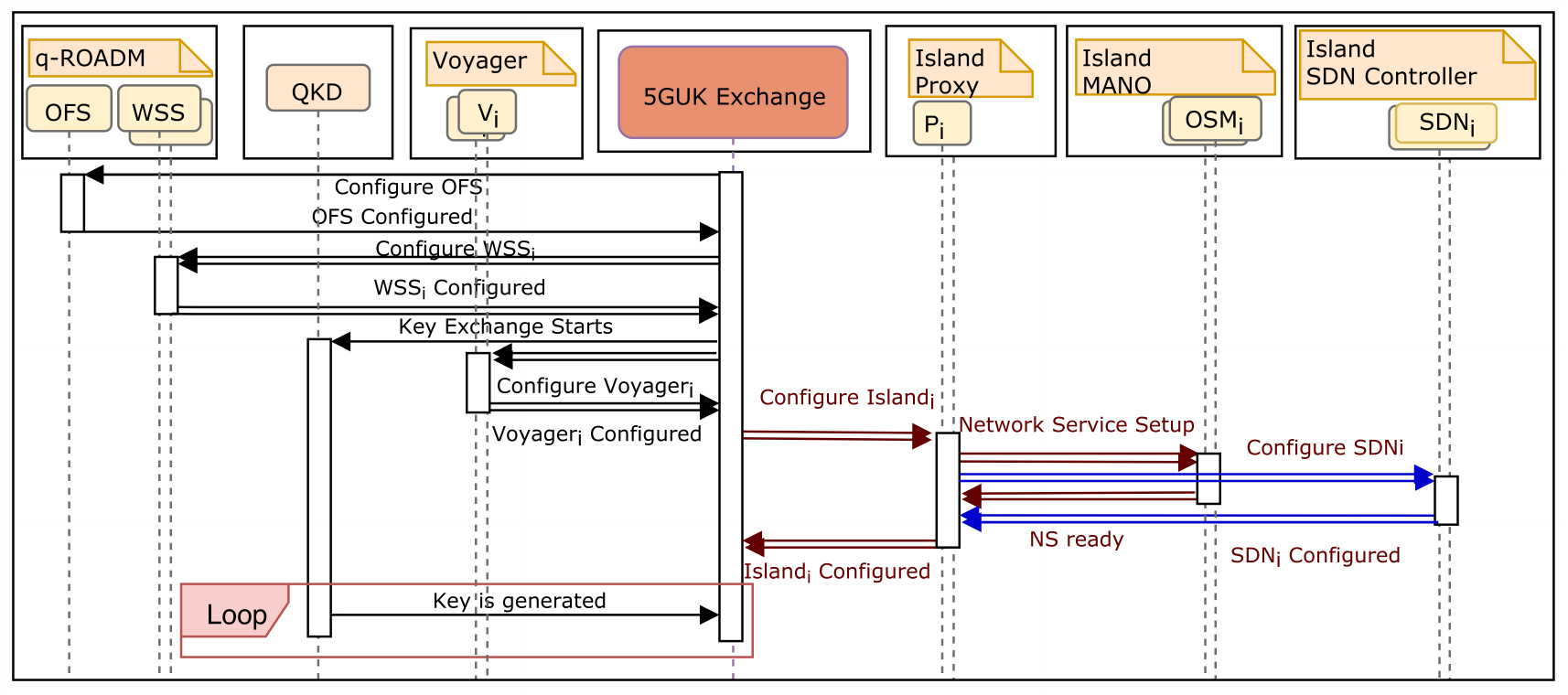}
\caption{Sequence diagram of Quantum-Secured Inter-Domain 5G Service Orchestration and On-Demand NFV Chaining.}
\label{fig:sequence_diagram}
\end{figure*}

Specifically, each NS deployed through 5GUK Exchange on each island is assigned a specific VLAN, where all VLANs are aggregated in a VLAN trunk through a layer 2 Corsa openflow switch.As an open reference design for open and disaggregated converged packet/optical switch, the voyager switch then convert packet stream into a lightpath (and vice versa) where 
coherent bandwidth variable transponder (with speed up to 200Gbps)  assigns each VLAN to a respective coherent link with controlled bit-rate and modulation format (QPSK, 8-QAM, 16-QAM) (see QIDCM in \Cref{sec:qidcm}). The {bandwidth variable transponder (BVT)} ports of the Voyager {depicted in~\Cref{fig:arch} of} each island are then multiplexed in a WSS (Finisar) and coupled to the island’s DV-QKD quantum channel generated by an IDQ Clavis 2. An additional server that interfaces the IDQ Clavis 2 with the QIDCM is also responsible for providing the quantum keys to software encryptors that encrypt NFV data. In this paper, ChaCha20 algorithm is used in the proposed software encryptor~\cite{bernstein2008chacha}. 

The multiplexed quantum and classical signals are transmitted over the same fiber in a coexisting scheme to the proposed fully functional low-loss, 4-degree q-ROADM where the mixed signals arriving from each island are first demultiplexed through a WSS and coupled to an optical space switch (Polatis), both controlled by QIDCM. 5 km fibres are deployed in the testbed to connect each island to {CD} q-ROADM. For the output ports of the q-ROADM, we have two options according to the destinations: 
\begin{enumerate*}[label=\arabic*), ref=\arabic*]
\item quantum and classical data channels are multiplexed again in the bypass output ports; and 
\item quantum and classical data channels are separated and dropped locally from the drop port.
\end{enumerate*}
On the bypass port of the q-ROADM, unamplified  quantum channels and amplified/filtered classical data channels are coupled through a 95:5 coupler. For the drop ports quantum and classical channels are separately driven to the island. On the islands receiver side, the mixed channels from the bypass ports are demultiplexed through a fixed arrayed waveguide grating (AWG). For both cases, total quantum channels losses over the q-ROADM do not exceed 6 dB, while worst case end-to-end power loss for the quantum channel connecting two islands via two bypass ports is around 10dB.



\section{5GUK Exchange: NFV MANO Layer for inter-island communication}
\label{Sec:5GExArch}

The connectivity and orchestration of VNFs are managed by the 5GUK Exchange (5GUKEx)~\cite{5GUKEx}. As per the scenarios mentioned in \Cref{sec:scenarios}, the NSes are interconnected to operate between the multiple islands. The 5GUKEx has been extended in this work to make it \emph{quantum} and \textit{Layer 0} aware. The 5GUKEx is composed of multiple specialized components supporting the inter-island network service composition and management. These include the NS broker to coordinate between multiple islands and an Inter Domain Connectivity Manager (IDCM) to create the underlying network connectivity. To complete the block, a thin layer of orchestration is hosted over each island called Island Proxy which communicates with the NS Broker to share the island capabilities and the NS catalogues.

{
\newcommand\figureSize{0.480}
\begin{figure*}
\centering
\begin{tabular}{cc}
\subfloat[\label{fig:AWG_Tempurature}] { \includegraphics[width=\figureSize\linewidth,trim={0 17 0 0},clip]{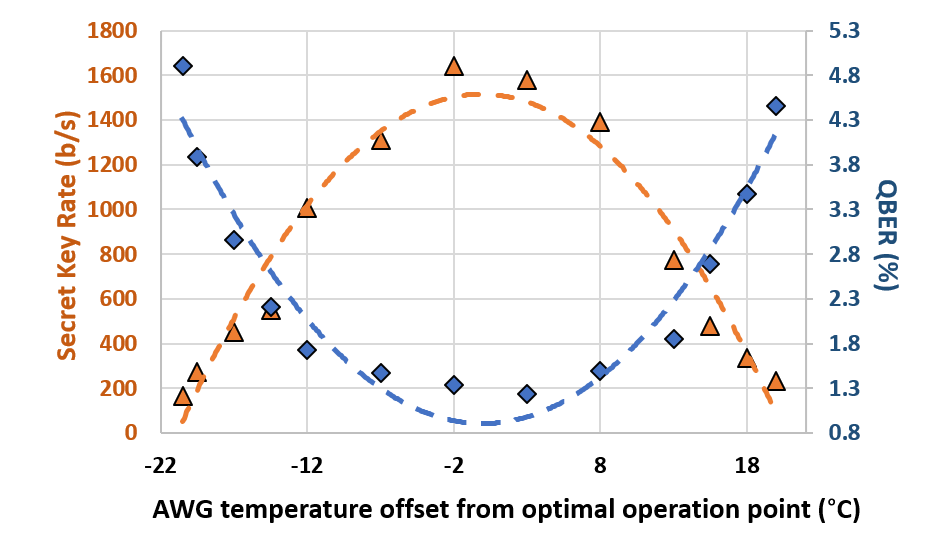} } & 
\subfloat[\label{fig:AWG_Profile}] { \includegraphics[width=\figureSize\linewidth,trim={0 2 0 0},clip]{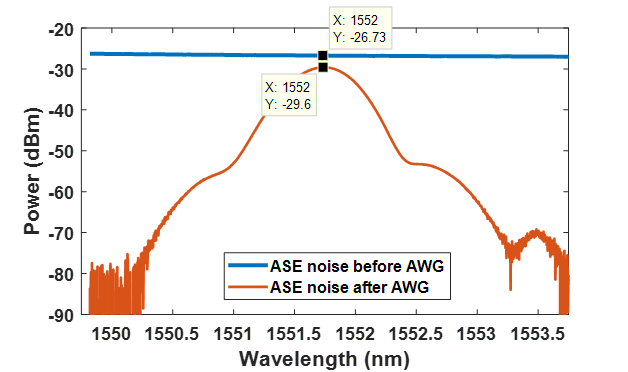} } \\
\end{tabular}
\caption{Arrayed waveguide grating performance characterisation for quantum channel. (a): Quantum secret key rate and QBER vs AWG temperature; (b): Arrayed waveguide grating filter profile for quantum channel}
\end{figure*}
}

\subsection{Quantum-aware Network Service Broker}
The Quantum-aware Network Service Broker (QNSB) interfaces between the Island Proxy and the 5GUKEx. It receives island registration requests containing the Network Service Descriptors (NSDs) and Virtual Network Function Descriptors (VNFDs) as part of the NS catalogues from the Island Proxies. It registers the island on 5GUKEx and passes this information to the Network Service Manager (NSM). The QNSB is invoked while instantiating, deploying and terminating the NSes across islands. It provides unique ID to each island during registration to track and monitor the deployed NSes.

\subsection{Network Service Manager}
The Network Service Manager (NSM) is responsible for the lifecycle of the deployed inter-island NS. It stores the NS catalogues shared by the islands during registration and provides them to the NS Composer for creating end-to-end NS. It also communicates with the QNSB using the south-bound APIs while managing the lifecycle of the inter-island NS.

\subsection{Quantum-Aware Inter-Domain Connectivity Manager}\label{sec:qidcm}
As described in \cite{5GUKEx}, the Quantum-aware IDCM is responsible of provisioning the inter-island network connectivity using the underlying SDN controller. For this experiment, the IDCM is extended to make it \textit{quantum} and \textit{layer 0} aware. For this particular use-case, in addition to the dynamic L2 network slicing, a block for quantum control and Quantum aware flex-grid Routing and Wavelength Assignment (QRWA) is created. As shown in \Cref{fig:arch}, the QKD control is designed to check if the inter-island NS is requested to be a secure end-to-end service. In this case, the QKD starts the Quantum Key Exchange between the two islands before setting up the L2 packet network. Additionally, the QRWA controls the configuration of the underlying q-ROADM composed of WSS and OFS. The QRWA configures the connecting ports and assign the required wavelengths to the configured ports
\color{black}

\subsection{Network Service Composer}
The Network Service Composer exposes the North Bound Interface (NBI) of 5GUKEx and hosts a Graphical User Interface allowing users to compose the inter-island NS by combining the network services shared by individual islands. The composition results in templates of inter-island NS that the user can choose to deploy. A deployment request invokes the NSM and NSB which interact with the Island Proxy to deploy an NS on individual islands.


\subsection{Island Proxy}
Each island is assumed to host its own ETSI NFV Management and Orchestration (NFV-MANO) system. The Island Proxy consumes the north-bound APIs of the NFV-MANO and registers island to the 5GUKEx using the security certificates shared with each island. On registration, Island Proxy shares the NS catalogue with the 5GUKEx. At a later stage, the island can either reconnect (without updating the NS catalogue) or re-register (sharing the updated NS catalogue). For this experiment, the Island Proxy, as described in~\cite{5GUKEx} has been extended to make it layer 0 aware using the NBIs of the underlying SDN controller at the individual islands. In addition to the creation of L2 flows, it configures the underlying voyager switches with the desired wavelengths and the VLANs. The Island Proxy communicates the VLAN and wavelength information to the NS Broker, which in turn provides the information to the QIDCM to create the inter-island network.

\Cref{fig:arch}. shows the quantum-secured 5GUKEx architecture and the experimental setup, including the physical optical network that supports coexistence and independent switching of quantum channels and classical data channels. For this experiment, we have connected four different islands to 5GUKEx, sharing multiple NSes with the 5GUKEx. It allows composition of NSes across islands and dynamic selection of the quantum secured channel as needed. Each island has its own representative 5G network with radio access and optical backhaul along with separate compute and network resources administered individually by each island.
\color{black}


\Cref{fig:sequence_diagram} shows the exchange of control messages between the various components of Quantum-Secured Inter-Domain 5G-Service Orchestration. For the sake of simplicity, the proceedings between 5GUK Exchange components are not included in the diagram as is already been discussed in this section. 

\begin{itemize}
\item 5GUK Exchange is the central point allowing users to  create the multi-domain inter-island NS (iNS). In additon to the functionality explained in this section, the iNS also define the wavelength and selects the virtual link to be secured using quantum channel. Before establishing the L2 virtual link between the NSes, the QIDCM configure the underlying OFS and WSS as a part of q-ROADM with appropriate wavelengths(for classical and quantum channels) and the port mappings. 

\item As shown in \Cref{fig:sequence_diagram}, once the OFS and WSS (in q-ROADM) is configured, the QIDCM contact the selected islands (for secure link) and starts key exchange between them.
\begin{itemize}

\item In case of unsecured channel, the QKD link is not established rather a classical optical channel is created.
\end{itemize}

\item Since, Key exchange is a long and asynchronous process, the 5GUKEx starts deploying the requested NSes and links between them. At first, \Cref{fig:sequence_diagram} shows the interaction of 5GUKEx to the individual islands where QNSB contact proxy on each side to configure Voyagers with appropriate parameters (modulation, freq, VLANs, wavelength etc.) 

\item On successfully configuring the voyagers, the QNSB instruct proxy to configure and deploy NSes on respective islands. The process is as explained previously in this section, where NFVO on each island is requested to deploy the VNFs and SDN controller is requested to create the appropriate flow rules to create an end-to-end iNS.

\item Finally, on establishment of the secured QKD channel, the data-path is established for iNS to be operational with the desired secured connection.
\end{itemize}

\color{black}

\section{Experimental Results And Analysis}
\label{sec:results}

{
\newcommand\figureSize{0.475}

\begin{figure*}[!t]
\centering
\begin{tabular}{cc}
\subfloat[\label{fig:Power_sweep_1ch_QPSK}] { \includegraphics[width=\figureSize\linewidth]{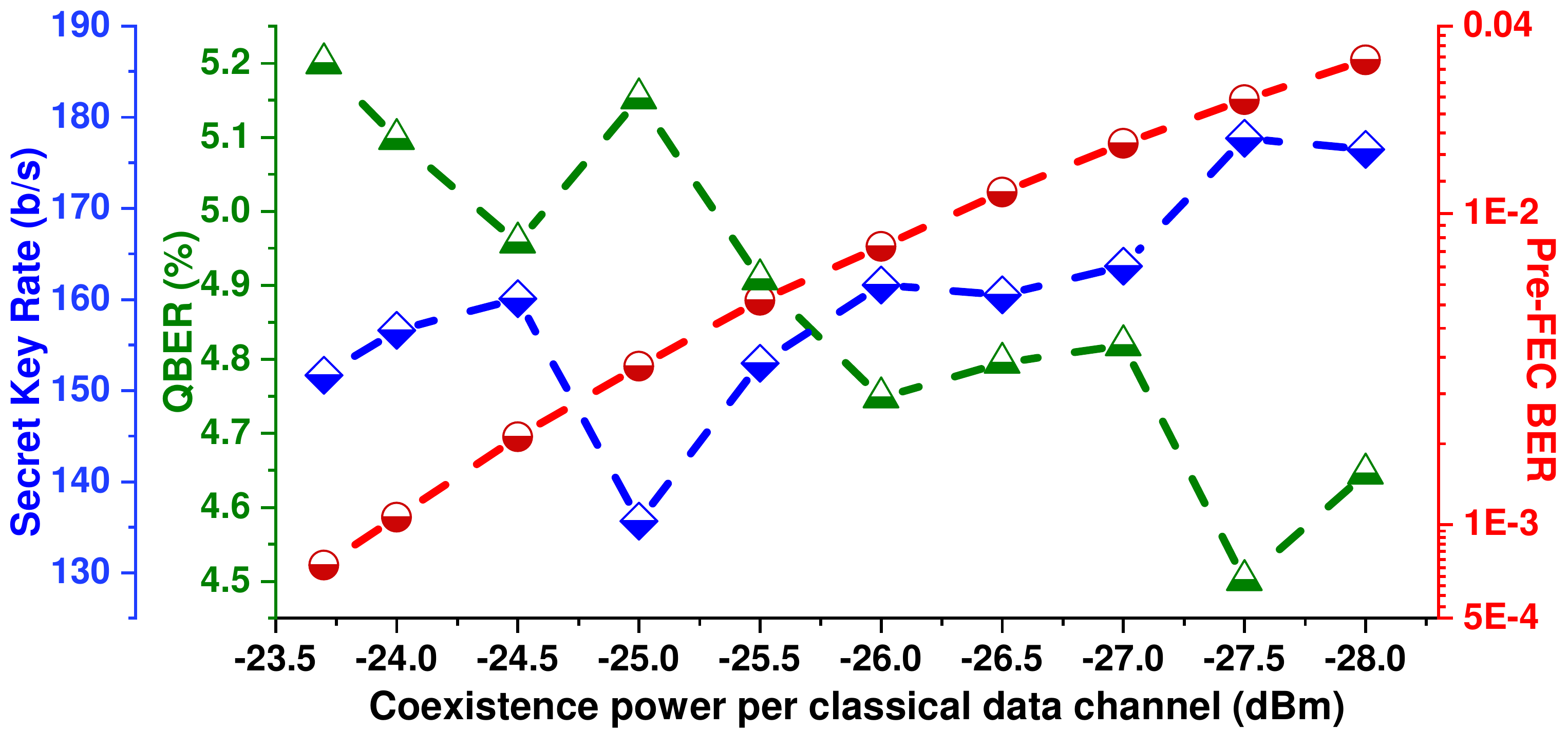} } & 
\subfloat[\label{fig:Power_sweep_2ch_QPSK}] { \includegraphics[width=\figureSize\linewidth]{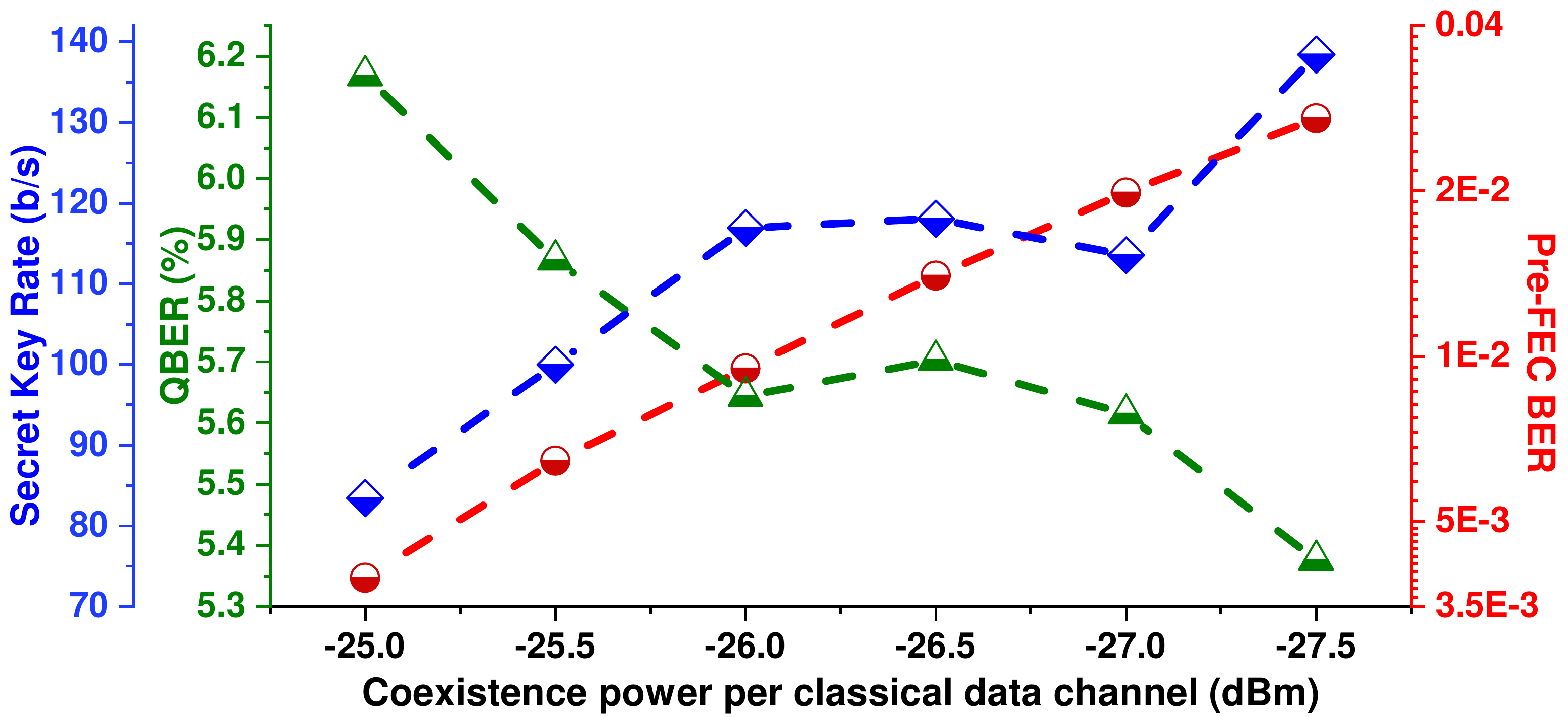} } \\

\subfloat[\label{fig:Power_sweep_3ch_QPSK}] { \includegraphics[width=\figureSize\linewidth]{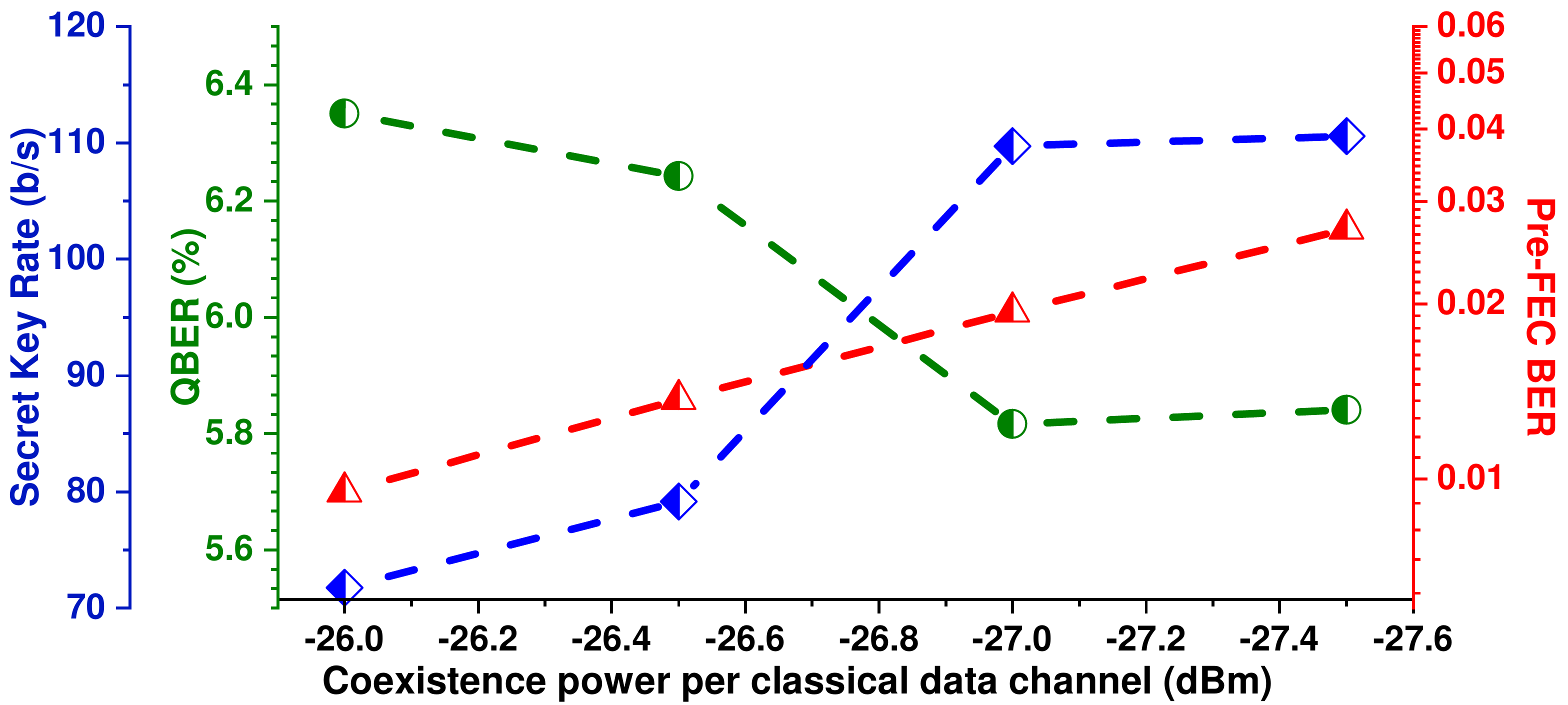} } & 
\subfloat[\label{fig:Power_sweep_1ch_16QAM}] { \includegraphics[width=\figureSize\linewidth]{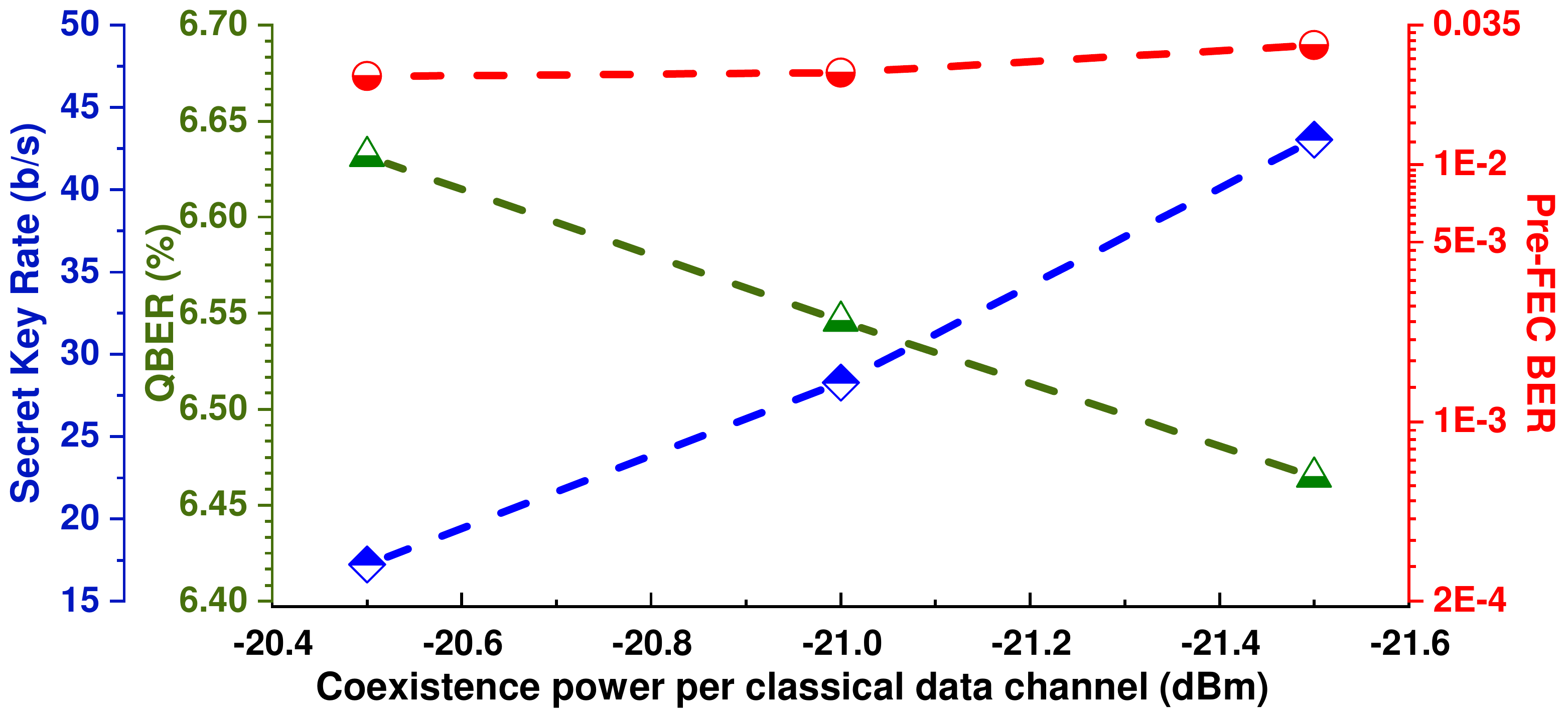} } \\
\end{tabular}

\caption{q-ROADM bypass port performance: QKD secret key rate, QBER and classical data channel pre-FEC BER vs per data channel power. (a) One data channel with PM-QPSK and 25\% FEC; (b) Two data channels with PM-QPSK and 25\% FEC; (c) Three data channels with PM-QPSK and 25\% FEC; (d) One data channel with PM-16QAM and 25\% FEC }\label{fig:Power_Modulation_Sweep_QKD}

\end{figure*}
}

\begin{figure}[hbtp]
    \centering
    \includegraphics[width=\columnwidth]{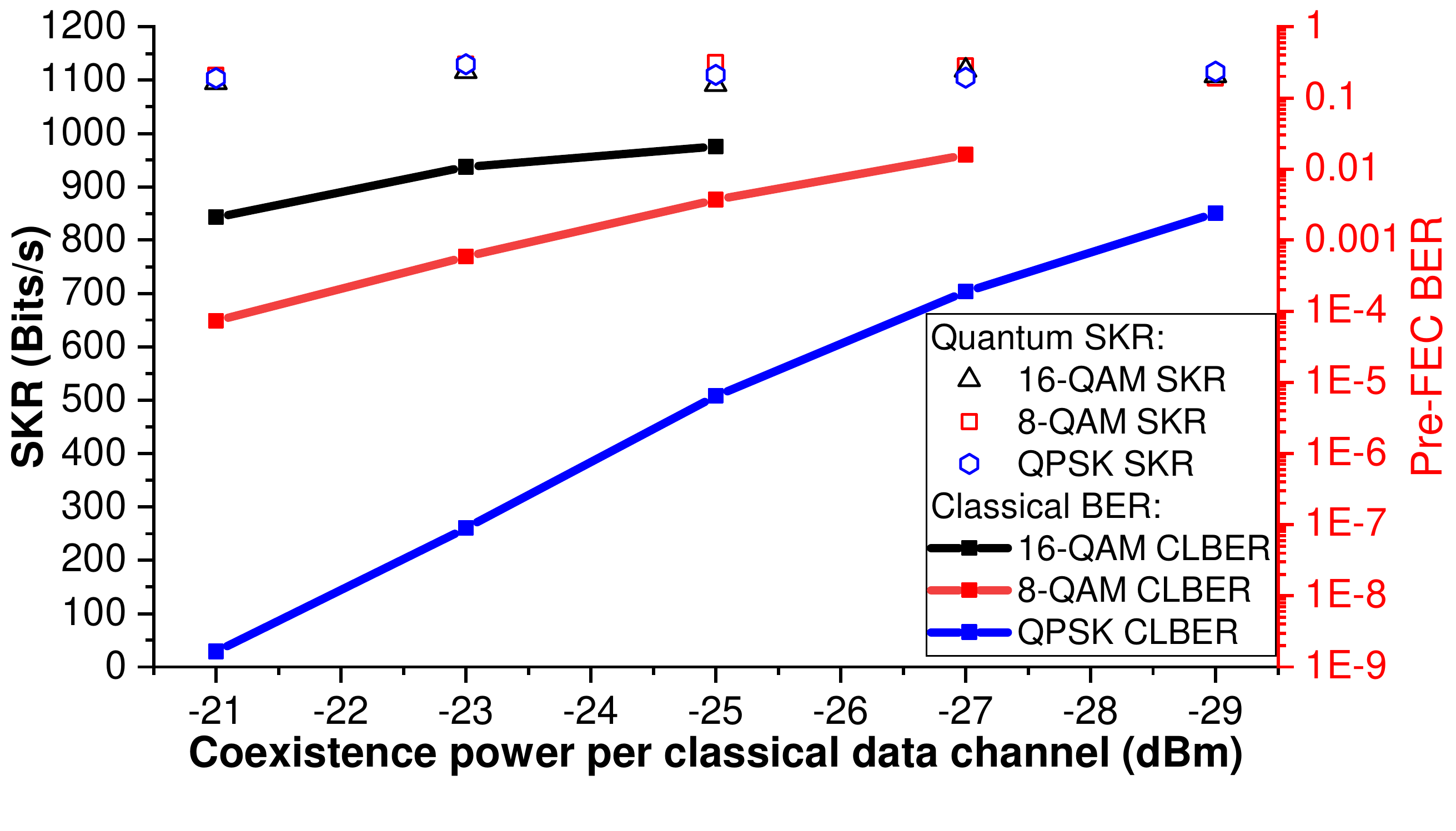}
    \caption{q-ROADM drop port performance: QKD secret key rate and classical data channel pre-FEC BER vs per data channel power}
    \label{fig:Modulation_Drop_Port_OFC_PDP}
\end{figure}

In this section, we first evaluate the performance of data plane and illustrate the results obtained from the proposed testbed architecture shown in \Cref{fig:arch} against various data channel coexistence power and modulation formats. The physical layer results are collected by establishing lightpaths 1) between islands both connected to the bypass ports of the q-ROADM (Island 1 and Island 3), and 2) one island connecting the bypass port of q-ROADM (Island 1) and the other one connects to drop port of q-ROADM (Island 4). Further, the control and management plane results for three dynamically switched scenarios as mentioned in \Cref{sec:scenarios} are presented. The timing of network components, computing resources, and quantum equipment are also depicted to demonstrate and confirm the flexibility and programmability of the proposed control and management plane framework over the physical infrastructure.

\subsection{Physical Layer Results} \label{sec:res:data}

In this subsection, the physical layer characteristics of the proposed testbed architecture shown in \Cref{fig:arch} such as classical data channel BER and quantum secret key date are investigated and the results are depicted. The arrayed waveguide grating in each island as illustrated in \Cref{fig:arch} is regarded as a static DEMUX to decouple the classical data channels and the QKD channel. The characteristics of the AWG can significantly affect the performance of quantum channel securing inter-islands NS. Therefore, central frequency of the AWG needs to adjusted and calibrated to the frequency of the quantum channel via temperature control to optimize QKD performance. Therefore, we test the performance of quantum channel in terms of secret key rate and  quantum bit error rate (QBER) against the increasing temperature of AWG by directly connecting Alice and Bob via AWG. In \Cref{fig:AWG_Tempurature}, it is illustrated that the quantum secret key rate increases while QBER decreases with the increasing AWG temperature until the best operational temperature, and then the performance of quantum channel degrades with further raising temperature of AWG.

At the optimal operational temperature AWG, the central frequency of AWG port for QKD becomes identical to the operational frequency of DV-QKD IDQ Clavis 2 system. In this case, it gives the minimum power loss to the quantum channel, thus achieving the highest quantum secret key rate and lowest QBER. To quantify the power loss of QKD channel passing through AWG, ASE noise used and fed into the port supporting QKD channel of the AWG in the characterisation experiment. Power spectral density of the ASE noise before and after injected into the AWG port are measured via optical spectrum analyzer. In \Cref{fig:AWG_Profile}, the filtering profile of the AWG port for the quantum channel is illustrated under the optimal operational temperature for the measurement. For the AWG used in the experiment, it gives around 2.9 dB power loss to the quantum channel after careful AWG temperature adjustment.
{
\newcommand\figureSize{0.30}
\begin{figure*}[htbp]
\centering
\begin{tabular}{ccc}
\subfloat[\label{fig:First_Scenario}] { \includegraphics[width=\figureSize\linewidth]{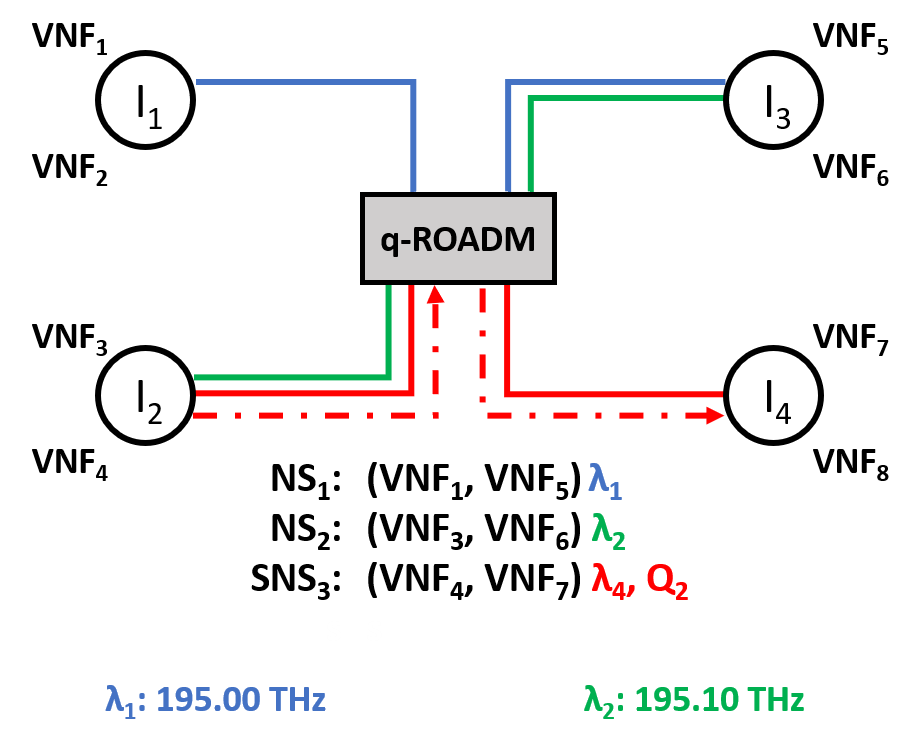} } & 
\subfloat[\label{fig:Second_Scenario}] { \includegraphics[width=\figureSize\linewidth]{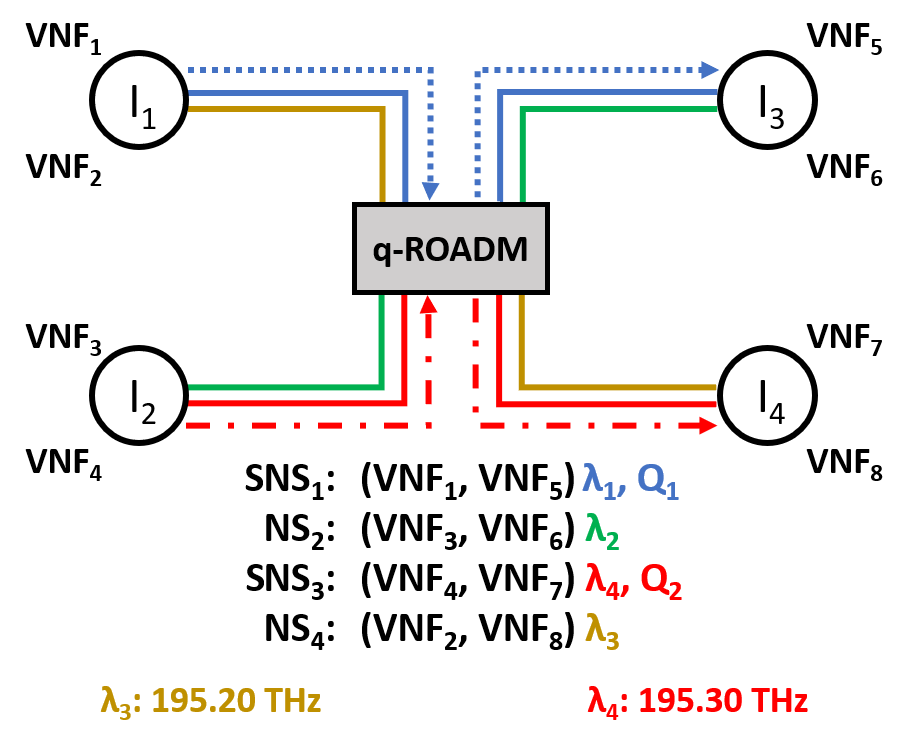} } &
\subfloat[\label{fig:Third_Scenario}] { \includegraphics[width=\figureSize\linewidth]{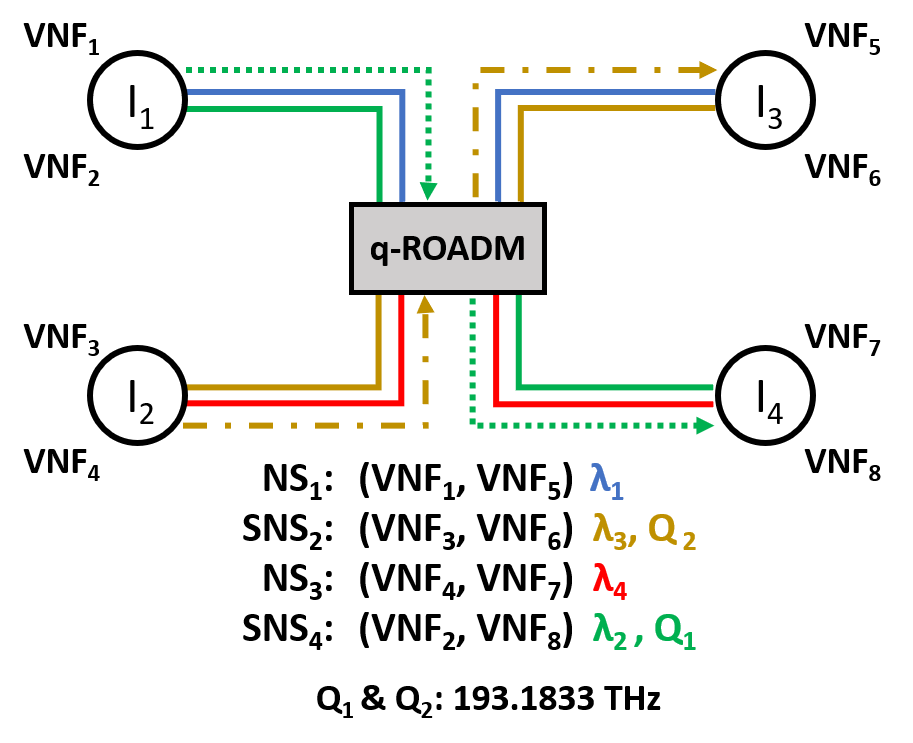} } \\
\subfloat[\label{fig:First_Scenario_Results}] { \includegraphics[width=\figureSize\linewidth]{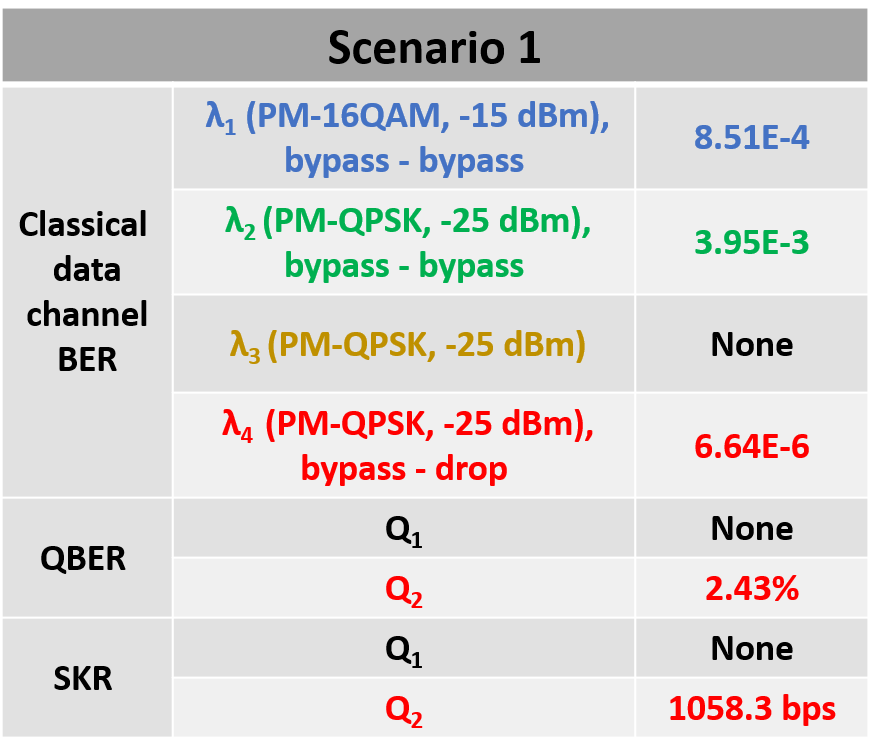} } & 
\subfloat[\label{fig:Second_Scenario_Results}] { \includegraphics[width=\figureSize\linewidth]{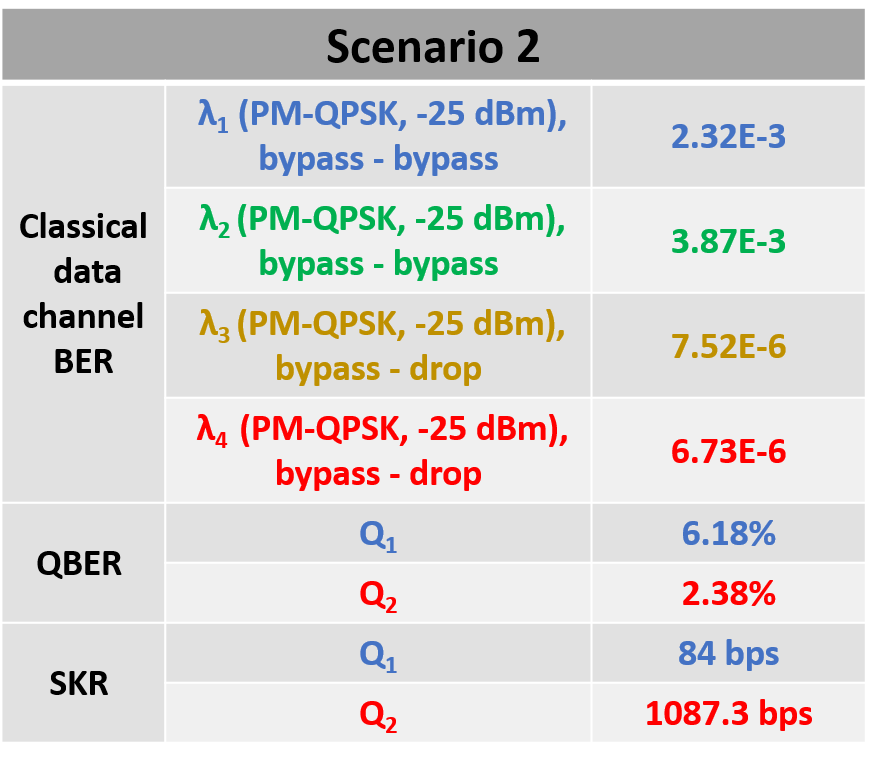} } &
\subfloat[\label{fig:Third_Scenario_Results}] { \includegraphics[width=\figureSize\linewidth]{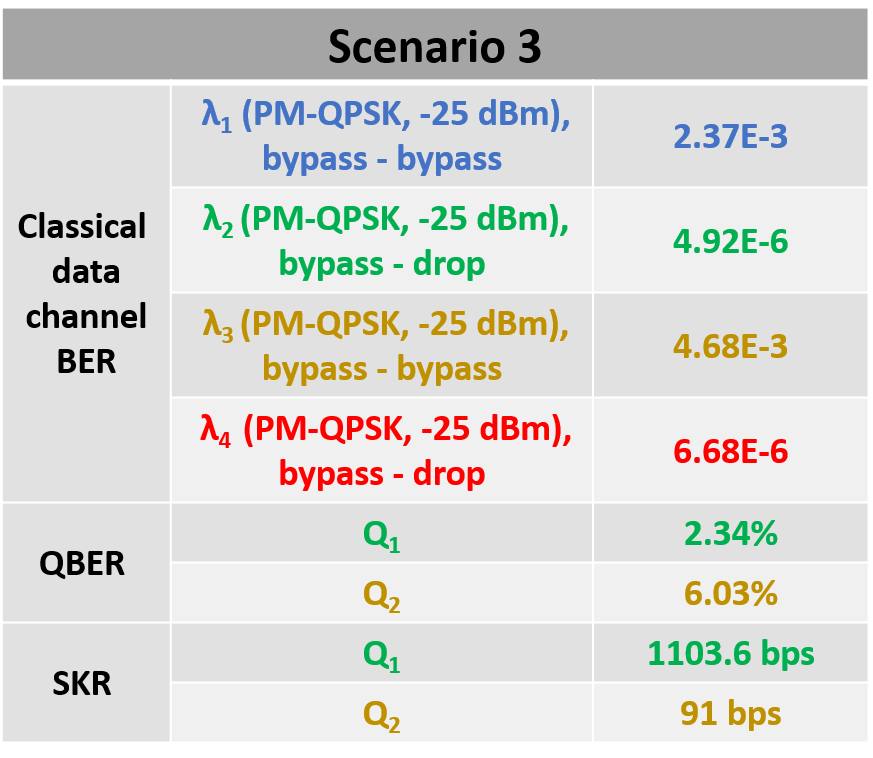} } 
\end{tabular}
\caption{Three dynamically switched experiment scenarios and their results. (a) The first scenario; (b) The second scenario; (c) The third scenario; (d) Results of the first scenario; (e) Results of the second scenario; (f) Results of the third scenario.}\label{fig:3_Scenarios}
\end{figure*}
}

To investigate the performance of the proposed quantum secured 5G service orchestration architecture over the flex-grid optical networks, we setup the experiment testbed as shown in \Cref{fig:arch}. It is depicted in \Cref{fig:Power_Modulation_Sweep_QKD} the physical layer characterization of classical and quantum channels coexistence scheme over the shared optical network between Island 1 and Island 3 via the bypass ports of the q-ROADM. The figure reveals the per channel coexistence power operational windows. The results show the performance of QKD coexisting with a single data channel (195.0 THz) using PM-QPSK and 25\% FEC overhead as shown in \Cref{fig:Power_sweep_1ch_QPSK}, two data channels (195.00 THz and 195.10 THz) both using PM-QPSK and 25\% FEC overhead as shown in \Cref{fig:Power_sweep_2ch_QPSK}, three data channels (195.00 THz, 195.10 THz and 195.20 THz) using PM-QPSK and 25\% FEC overhead as shown in \Cref{fig:Power_sweep_3ch_QPSK}, and PM-16QAM signal of a single data channel (195.00 THz) with 25\% FEC overhead as shown in \Cref{fig:Power_sweep_1ch_16QAM}, in order to sustain both the classical and quantum channels with secret key rate (SKR) $>$ 0 and pre-FEC BER~$<$~threshold. 

To evaluate the performance of the proposed architecture, the per data channel power is varied. As a result, the overall aggregated coexistence data channel power can be calculated as: aggregated power = per channel coexisting power + $log_{10}$ (number of channels). A maximum of 178 bps secret key rate for QKD can be achieved on the bypass channel for one coexisting classical data channel with a quantum channel at -28 dBm coexistence power. The SKR decreases by 27\% when increasing the coexisting power to -25 dBm. For the case of two and three classical data channels, slightly worse performance with 138 bps and 110 bps SKR are achieved at -27.5 dBm per data channel coexistence power (aggregated power of -24.5 dBm and -22.73 dBm) respectively. The SKR drops approximate 15\% and 36\% respectively when the coexisting power level {increases} to -26 dBm compared to -27.5 dBm coexisting power.
As shown in \Cref{fig:Power_sweep_1ch_QPSK,fig:Power_sweep_2ch_QPSK,fig:Power_sweep_3ch_QPSK,fig:Power_sweep_1ch_16QAM}, the QBER drops while the quantum secret key rate raises against the decreasing power of coexistence data channels for all four scenarios. There are two main factors to limit the power level of data channel coexisted with the quantum channel in the proposed architecture: \begin{enumerate*}[label=\arabic*), ref=\arabic*]\item nonlinear impairments and \item crosstalk.\end{enumerate*} Higher data channel power would impair the quantum channel by introducing and creating higher level of Raman scattering \cite{da2014impact}. Moreover, the AWG within the local island to decouple the classical data channel and quantum channel cannot completely separate them, thus introducing crosstalk due to data channels to be received by quantum receiver Bob. For the classical data channels with lower coexisting power, the EDFAs shown in the system setup in \Cref{fig:arch} introduce higher penalty to them. As a result, it leads to higher pre-FEC BER for all the cases.


\begin{figure*}[htbp]
\centering
\includegraphics[width=\linewidth]{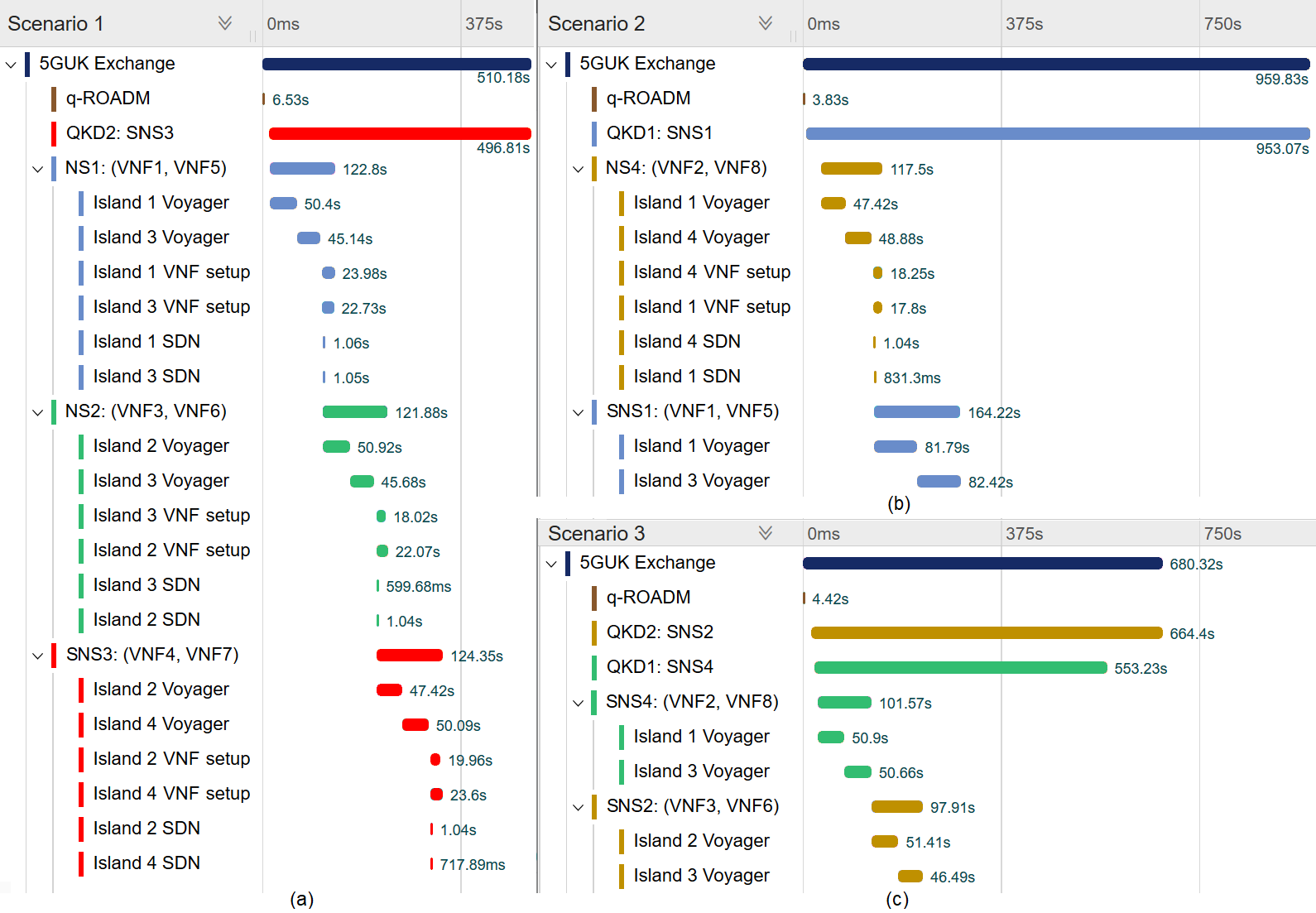}
\caption{Control plane and orchestration timings results for the three scenarios. (a) Scenario 1; (b) Scenario 2; (c) Scenario 3}
\label{fig:control-timings}
\end{figure*}

We also investigate the effect of modulation formats of the classical data channel over the quantum channel via the bypass port (between Island 1 and Island 3) and the drop port of the q-ROADM (between Island 1 and Island 4), as shown in \Cref{fig:Power_sweep_1ch_16QAM} and \Cref{fig:Modulation_Drop_Port_OFC_PDP} respectively. For bypass - bypass connection, A higher coexistence power is required when the data channel utilises PM-16QAM as modulation format, as compared to using PM-QPSK. As a result, it leads to higher level of nonlinear impairments interference and the crosstalk to the quantum channel for the islands communicating through the bypass ports of q-ROADM. \Cref{fig:Power_sweep_1ch_16QAM} shows that the minimum operational coexistence power for PM-16QAM is -21.4 dBm, with around 1 dB operational power window, while the PM-QPSK allows a broader, around 5 dB operational power range, as shown in \Cref{fig:Power_sweep_1ch_QPSK}. Considering the physical layer performance between Island 1 and Island 4 via drop port of q-ROADM, the shorter length fibre compared to fibre between Island 1 and Island 3 leads to lower power loss and less nonlinear impairments interference. The band-pass filter at the drop port placing before Bob is able to filter out-band crosstalk from WSS. Therefore, maximally 1100 bps quantum secret key rate can be achieved for various modulation formats and coexistence data channel power, as depicted in \Cref{fig:Modulation_Drop_Port_OFC_PDP}. Due to different noise level requirements, the pre-FEC BER of channels vary with their assigned modulation formats. Similar to \Cref{fig:Power_Modulation_Sweep_QKD}, the pre-FEC BER of classical channels grows against the decreasing coexisting power level.



\subsection{Orchestrator And Control Plane Workflow And Results} \label{sec:res:Control_data}

The goal of this experiment is to present the flexibility of the proposed data plane architecture and the proposed control/ management plane framework. As described in \cref{sec:scenarios}, three test scenarios are devised to verify our testbed capabilities and configuration times during the establishment of new network services and the reconfiguration of existing ones.

In each scenario, one or more NSes are established and/or modified. Each NS is defined by its composing VNFs, latency, bandwidth, and security requirements, as well as its services starting time and Time To Live (TTL). In our experiments, each NS is composed of two chained VNFs (one NS with single VNF in each island), which can be secured by a QKD Clavis pair, shown as a Secured NS (SNS). To demonstrate the chaining of distributed VNFs over an optical network, the VNFs belonging to the same NS are hosted in different islands, interconnected by classical data channels and an optional quantum channel. The related optical parameters (i.e., modulation format and launch power) are defined according to the required quality of service of each NS. Up to four wavelengths are utilised to provide bidirectional connection for NS across multiple islands through the q-ROADM, as shown in \Cref{fig:3_Scenarios}. The frequencies used in our experiment are $\lambda_1$ 195.00 THz, $\lambda_2$ 195.10 THz, $\lambda_3$ 195.20 THz, and $\lambda_4$ 195.30 THz, with 100 GHz spacing to suit the grid of AWG deployed in the testbed.

The first scenario, as depicted in \Cref{fig:First_Scenario}, consists of three NSes with one being secured.
Next, we transit to the second scenario, \Cref{fig:Second_Scenario}, by dynamically adding an extra NS and by securing one of the existing NS with a second QKD pair, thus increasing the total number of NS to four with two secured.
Finally, in the third scenario, \Cref{fig:Third_Scenario}, the wavelengths between two NS are swapped and the two quantum channels are reconfigured to secure two other NS. 
\Cref{fig:First_Scenario_Results,fig:Second_Scenario_Results,fig:Third_Scenario_Results} shows the data collected regarding the classical channels BER and the quantum SKR and QBER on all the scenarios.


\Cref{fig:control-timings} shows the timings obtained for management, control, and data plane functions for the three scenarios. 
\Cref{fig:control-timings} (a) details the timings involved in the configuration of the first scenario. As described by \Cref{fig:sequence_diagram}, the process starts with the 5GUK Exchange orchestration service configuring the q-ROADM, followed by the QKD system(s). The q-ROADM configuration time is the sum of the configuration time of the OFS and the WSSs. The time of OFS is relatively constant. Thus, major variations on the total q-ROADM setup time are due to the individual WSS, which varies upon the complexity of configurations required. 

After the optical paths successfully setting up, the orchestration platform triggers the establishment of the quantum channel immediately, as this phase requires a significantly longer period than any other steps. The QKD time comprises the start of the encryption software, the configuration of the IDQ Clavis equipment, and the waiting time for the first confirmation that the quantum keys are being generated successfully. Although key generation may start before this confirmation message, our software waits for the acknowledgment as a reliability measure. The QKD initialization time mainly depends  on  the  attenuation  of  the  quantum  channel  where the longer time is associated with the higher power attenuation.


Meanwhile, in parallel to the QKD establishment, the 5GUK Exchange service proceeds to the setup of the NSes. First, the Voyager in each island is configured, taking from 45s to 55s which involves wavelength and power modification, and up to 80s - 90s when also changing the modulation format. Since NS\textsubscript{1} does not coexist with any quantum channel, it has an extra power budget, allowing it to be configured to a denser PM-16QAM modulation scheme. As shown in \Cref{fig:First_Scenario_Results}, the channel ($\lambda_4$) dropped on the drop port performs better BER wise when compared with the channels passing through the bypass port,as revealed in \Cref{sec:res:data}.
Afterwards, as described in \Cref{Sec:5GExArch}, the QNSB requests the island proxies to deploy the requested NSes. In our results, we have considered the VNF deployment and activation time. Since we have used the same NS and VNFs on each island and the islands reside in the same server rack, the NS deployment and activation times are similar for each island. Finally, the island proxy communicates with the local SDN controller to create the L2 flow rules to establish the data plane.

\Cref{fig:control-timings} (b) shows the transition timings from the first to the second scenario. Following the same logic of the first scenario, the q-ROADM is reconfigured first, followed by the setup of a second {QKD} pair, used to secure the previously created NS1. Then, a new NS is created (NS\textsubscript{4}), following the same process described earlier. After provisioning the fourth classical data channel and the second quantum channel, the power budget of NS\textsubscript{1} ($\lambda_1$) is not enough to allow the use of PM-16QAM as modulation scheme. Therefore, the modulation format NS\textsubscript{1} is dynamically switched to the PM-QPSK modulation, hence allowing a lower coexistence power, -25 dBm instead of -15 dBm. When comparing the timing results of the first and the second scenarios, the main differences are QKD and the islands 1 and 3 Voyager switch configuration time. 
The setup of QKD from Island 1 to Island 3 takes $\approx90$\% longer than the setup of QKD from Island 2 to Island 4 due to the higher end-to-end power attenuation associated the lightpaths passing through two bypass ports when compared to the bypass-drop port connection.
Also, an additional modulation format change operation explains the Voyagers slower configurations ($\approx82s$). 

Finally, \Cref{fig:control-timings} (c) shows the timings of the operations moving from scenario 2 to scenario 3. As described in \cref{sec:scenarios}, this scenario is created to prove the dynamicity of the 5GUK Exchange. On analyzing the control plane results, it is noticeable that the configuration of the NS only includes Voyager switch configuration, which shows that it is possible to change the optical layer configuration without redeploying the VNFs and the local island networking, hence avoiding further configuration of the SDN controller and the NS setup layer. Lastly, the measured difference in the QKD time follows the same behaviour from scenarios one and two. Due to the difference in attenuation, the QKD securing SNS\textsubscript{4} between Island 1 and 4 (connected to the drop port) is established $\approx1m$ $30s$ faster than the QKD securing SNS\textsubscript{2} between Islands 2 and 3 with both connected to the bypass port of q-ROADM.

\section{Conclusion}
\label{sec:conclusion}
In this paper, we have proposed a novel AoD based colourless, directionless and contentionless QKD switch-enabled flex-grid ROADM architecture. The performance of the proposed q-ROADM framework has been experimentally investigated and prove to be a solid architecture to enable classical data channels switching and quantum key switching simultaneously. We have experimentally demonstrated on-demand composition of quantum secured network services for the first time by chaining distributed VNFs over an optical network employing a q-ROADM. The control plane results show the flexibility of the proposed solution, allowing dynamic establishment of new network services and reconfiguration of the existing ones.

\section*{Acknowledgement}
This work was funded by EU funded projects UNIQORN (820474), Metro-Haul (761727), and UK EPSRC funded projects: TOUCAN EP/L020009/1, INITIATE EP/P003974/1, UK Quantum Hub for Quantum Communications Technologies EP/M013472/1.

\bibliographystyle{IEEEtran}
\bibliography{IEEEabrv,refs}

\end{document}